# Magnetic Compasses and Chinese Architectures

**Amelia Carolina Sparavigna**

Politecnico di Torino

**Abstract:** In this paper, we are discussing the effect of the use of magnetic compasses on the orientation of ancient Chinese architectonical complexes. As Czech researchers proposed in 2011, assuming these complexes ideally oriented by the ancient architects along north-south direction, in the case that the surveys were made by means of magnetic compasses, we can find the axes of the complexes deviating from the cardinal direction, according to the local magnetic declination that existed at the time the structures were built. Following this idea, here we discuss some examples of possible alignments obtained by means of magnetic compasses, concluding that the Chinese surveyors adopted, during the Ming dynasty, a method based on the magnetic compasses. The architects of the antecedent Yuan Dynasty probably used an astronomical method.

An axis exists, known by the Latins as the "axis mundi" (the cosmic axis), about which the vault of the Heaven seems rotating. This axis, projected on the Earth surface, is giving the cardinal north-south direction. Some ancient monuments show that this axis was considered as very important. The Egyptian pyramids of Giza, for instance, were remarkably aligned along it by the ancient surveyors that used the rising and setting of stars or the shadow of a gnomon to find the true North [1]. The same is true for the Elamite complex of the Dur Untash ziggurat that has the directions of the diagonals less than a degree different from the cardinal directions [2]. Anyway, it is in China that the orientation of building and monuments according to the "axis mundi" became a rule.

As explained in [3], it was from the Neolithic times that the settlements were oriented along the north-south direction, in agreement with a solar principle of south-facing entry, prodromal of the theory of Qi, the vital energy that flows in any living thing, and of the Feng Shui geomancy. During the Xia and Shang dynasties, the palaces became a symbolic representation of the cosmos, based on a square layout, oriented strictly north-south, "since Qi flows that direction" [3]. Today, the north-south orientation is considered one of the general characteristics of the Chinese imperial urbanism [4], that we can easily see in the regular subdivision of the urban sites and consequently in the orientation of the buildings. As we stressed in [5], also Francis John Haverfield (1850-1919), British historian and archaeologist, noted the regular, cardinally oriented, urban planning of the ancient Chinese towns. In his book of 1913 [6], he proposed that these towns were laid out in such a manner according to a very old agrarian system. Actually, the Chinese urban planning is also including symbolisms concerning cosmology, geomancy, astrology and numerology [3,4], to have locally the harmony and balance observed in the heavens. And, as we will see in the following discussion, the Chinese architecture is also linked, in some cases, to the Earth's magnet field.

The phenomenon of magnetism was known since ancient times because of the existence of lodestones. After the observations that oblong objects made of lodestone were attracted by the poles of the Earth, some methods for determining the direction towards them were invented. It was in China that the first magnetic compasses were used, during the ancient Han dynasty, the second imperial dynasty ruling from 206 BC–220 AD [7]. Actually, they were used for divination and had the shape of a spoon, as shown by the model in the Figure 1 [8]. They were not used as compasses for orientation; at the time of the engineer and inventor Ma Jun (c.200–265 AD), the south-pointing chariots were preferred. These chariots were mechanisms not employing the magnetism [9]. As told in [8], the first mention of a spoon made of lodestone and observed pointing in a cardinal direction is in a Chinese work composed between 70 and 80 AD. "But when the south pointing spoon is thrown upon the ground, it comes to rest pointing at the south"[10].

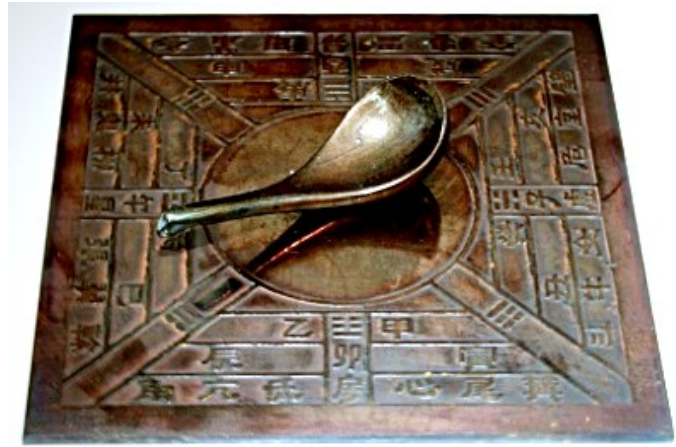

**Figure 1:** Model of an ancient Chinese compass having the shape of a spoon (Courtesy Wikipedia).

As explained in [11], in 1044 the Collection of the Most Important Military Techniques recorded that some magnetized fish-shaped objects were placed in a water-filled bowl, enclosed by a box, and were used to determine the direction, alongside the south-pointing chariot. However, "it was not until the time of Shen Kuo" that the earliest magnetic compasses were for navigation [11]. In his written works, Shen Kuo (1031-1095), a Han Chinese scientist and public official of the Song dynasty, made the first known explicit reference to the magnetic compass-needle and the concept of true North [11]. Actually Shen Kuo wrote, "the needle will point south but with a deviation", and also stated that "[the magnetic needles] are always displaced slightly east rather than pointing due south" [11]. Let us note that Shen Kuo was using an orientation due South, and that, since the needle was displaced slightly east, he was observing, in modern terms, a negative declination (the magnetic declination is the angle on the horizontal plane between magnetic North and true North).

Compasses began to appear around 1300 in Medieval Europe and Islamic world. Of the first European compasses, we have a description given by Petrus Peregrinus of Maricourt, a 13th-century French scholar and engineer, who wrote what we can consider as the first extant European treatise on magnetism. This treatise is in the form of a letter, probably composed in 1269. Peregrinus' letter consists of two parts. The first is discussing the properties of magnets, describing also the methods for determining their north and south poles. The second part of the letter describes some instruments that utilize the properties of magnets, ending with the Peregrinus' art of making a wheel of perpetual motion [12].

It is clear therefore that, during the medieval period, some very remarkable studies on magnetism existed, in particular those of Shen Kuo on the magnetic declination. Thereafter, the Chinese surveyors were aware of the existence of a difference between the magnetic and the astronomic north-south direction. If we assume, as told in [3], that Qi is the energy flowing this direction, because of the observed difference, a possible consequence was that of using the compass needle as the best instrument for determining the direction of this flow, because Qi was flowing through the magnetic material. Therefore, if we assume that Qi was considered in the orientation of buildings and architectonical complexes, it is plausible that we can find evidences in the layout of the Chinese architectures of a related surveying made according to the local magnetic field.

As proposed by the Czech researchers authors of Refs. 13 and 14, these evidences exist. "The tombs (pyramids) near the former Chinese capital cities of Xi'an and Luoyang ... show strong spatial orientations, sometimes along a basic South-North axis (relative to the geographic pole), but usually with deviations of several degrees to the East or West" [13]. For the authors of [13], the ancient Chinese surveyors used a compass, the needle of which "was directed towards the actual magnetic pole at the time of construction" [13]. By the way, let us remember that the magnetic poles of the Earth move with respect to the geographic poles. Consequently, the magnetic declination, besides depending on the location, is also changing over time.

In [13], the authors discussed the Chinese Pyramids, of which in [15] we considered a possible solar orientation. This use of magnetic compasses for the alignment of so old monuments could be questionable. However, in [13], we find mentioned the case of a remarkable architectonical complex; it is the Forbidden City in Beijing, dated from 1406 to 1420 AD (Figure 2). The layout of the Forbidden City was determined by means of a compass "as described (confirmed by some details), e.g. in the correspondence of Karel Slavíček, a Czech astronomer at Chinese imperial court in Beijing" [13]. Besides being evidenced by the historical report, the surveying of the complex by means of a magnetic compass is also in the proper period, that is, after the observations of Shen Kuo on the magnetic declination.

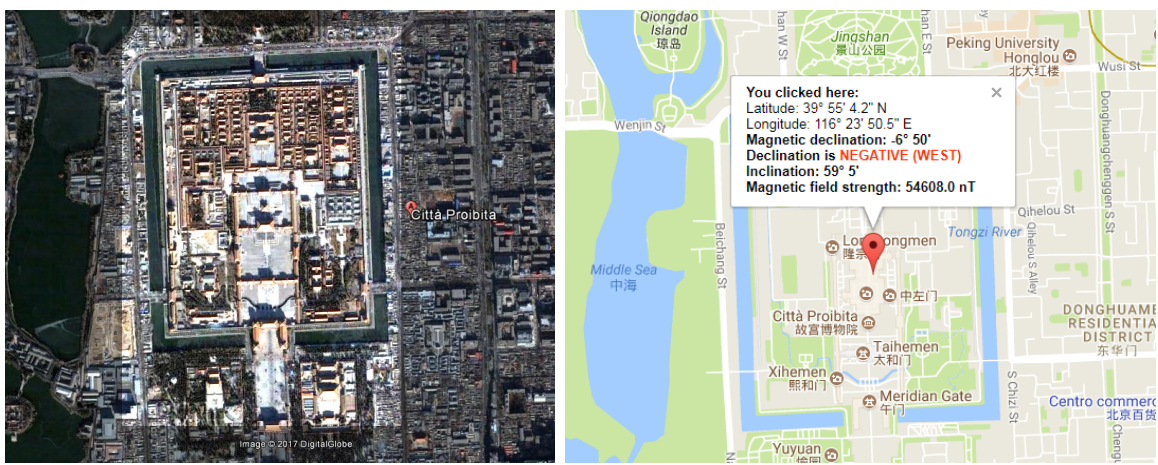

**Figure 2:** The Forbidden City of Beijing in a Google Earth image (on the left). On the right, a snapshot of the magnetic data given by the site http://www.magnetic-declination.com.

Let us consider the Forbidden City in the Google Earth images (Figure 2). We have a time series giving several images of the complex and therefore we can estimate a deviation from true North of (-2 ± 0.15) degrees (negative declination) at the time of its building. As told in [13], this is evident from historical records.

Today, the magnetic declination in Beijing is of 6°50' (negative). A web site is giving us this information at http://www.magnetic-declination.com. Here the snapshot of the result in the Figure 2.

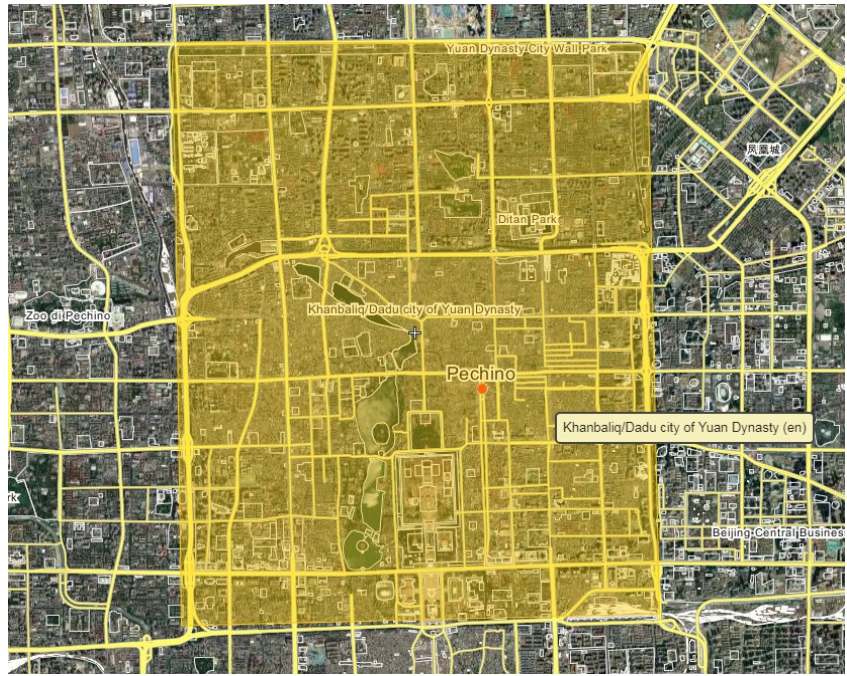

**Figure 3:** The Forbidden City inside Khanbalik, the capital of the Yuan Dynasty founded by Kublai Khan (Courtesy Wikimapia).

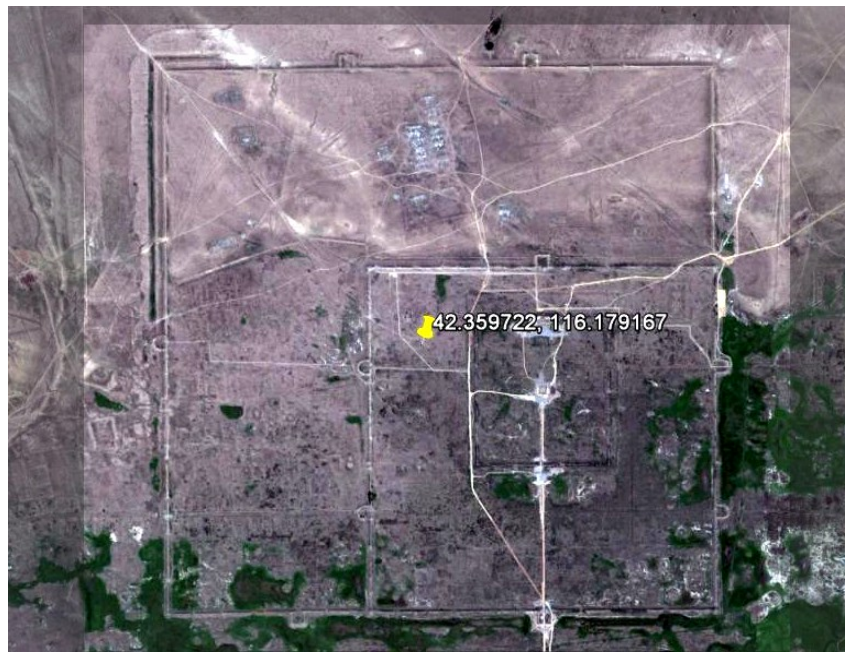

**Figure 4:** Xanadu (Courtesy Google Earth).

The Forbidden City was created inside the city of Khanbalik or Dadu, the "great capital" of the Yuan Dynasty founded by Kublai Khan (Figure 3) [5]. The architect and planner of Dadu was Liu Bingzhong, who was also the architect of Xanadu, the summer capital of Kublai Khan's empire [16]. The construction of the walls of Dadu began in 1264, while the imperial palace was built from 1274 onwards. The design of Dadu was based on a grid of 9 vertical axes and 9 horizontal axes, and followed the rules of "palaces in the front, markets in the rear", "left ancestral worship, right god worship" were taken into consideration [17].

We could ask ourselves if Liu Bingzhong used an astronomical or a magnetic surveying method for Dadu. Of course, it is difficult to deduce it from the streets of that part of the modern Beijing, corresponding to the ancient capital of the Yuan Dynasty. It seems that the streets have a better alignment along the cardinal directions than that observed for the Forbidden City. However, we have the possibility to investigate the layout of Xanadu that was built by the same architect. As we can see from the satellite image in the Figure 4, Xanadu is perfectly aligned along north-south direction. It means that the architect used a survey based of astronomical observations, not on magnetic compasses; of course, this is true if we consider the magnetic declination in Xanadu at the time of its foundation different from zero.

Let us assume the possibility that the architects of Yuan Dynasty preferred a survey based on astronomical observations, such as the rising and setting of stars. We have to consider that the Forbidden City was built by an emperor, Yongle (1360-1424), of another dynasty, the Ming dynasty, which was ruling China after the collapse of the Mongol-led Yuan dynasty. It seems therefore that the new rulers changed also the methods of land surveying, preferring the use of the magnetic compasses, probably because the magnetic compasses were evidencing the flow of Qi.

Another important architectonical complex that was built in Beijing in the same period reinforces this conclusion (this complex is not mentioned in [13]). It is the Temple of Heaven and its close Circular Mound Altar, which is an outdoor empty circular platform made of marble stones (Figure 5). These are the sites where the emperors of the Ming and Qing dynasties offered sacrifice to heaven and prayed for harvests. Located south of the Forbidden City, the original temple was completed together with the Forbidden City in 1420 during the reign of the Yongle Emperor. From the Figure 5, we can see that we have again a deviation of about 2 degrees, congruent with the use of a magnetic compass for surveying the site. Probably, the surveyors of the Forbidden City and of the Temple of Heaven were the same or used the same method.

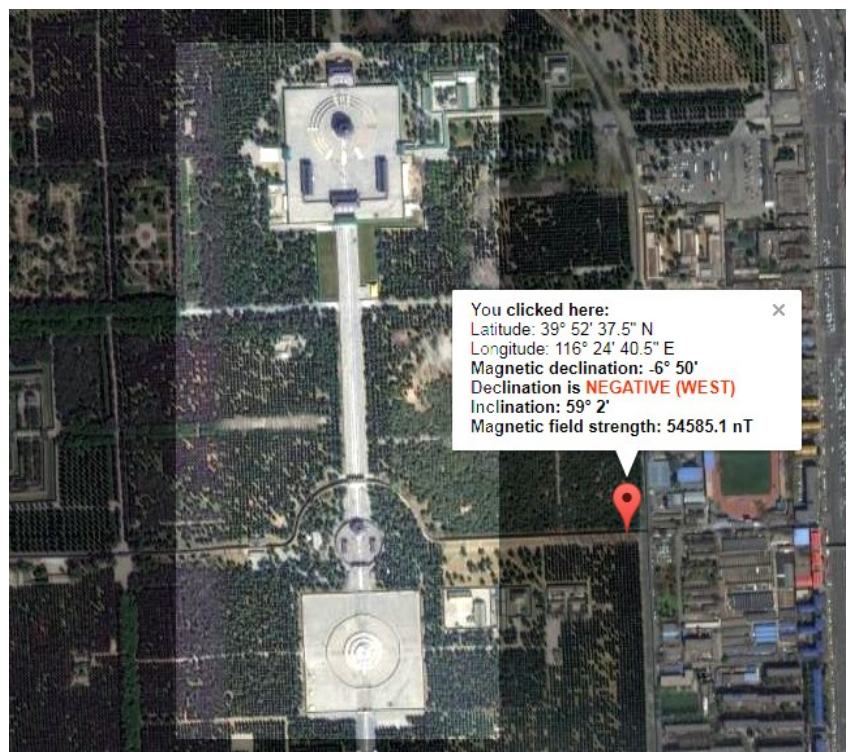

**Figure 5:** Snapshot of the result given by the site http://www.magnetic-declination.com for the Temple of Heaven in Beijing.

We have to tell that there are other architectonical complexes in Beijing displaying larger angles. One is the Temple of the Earth shown in the Figure 6. The angle is of six degrees.

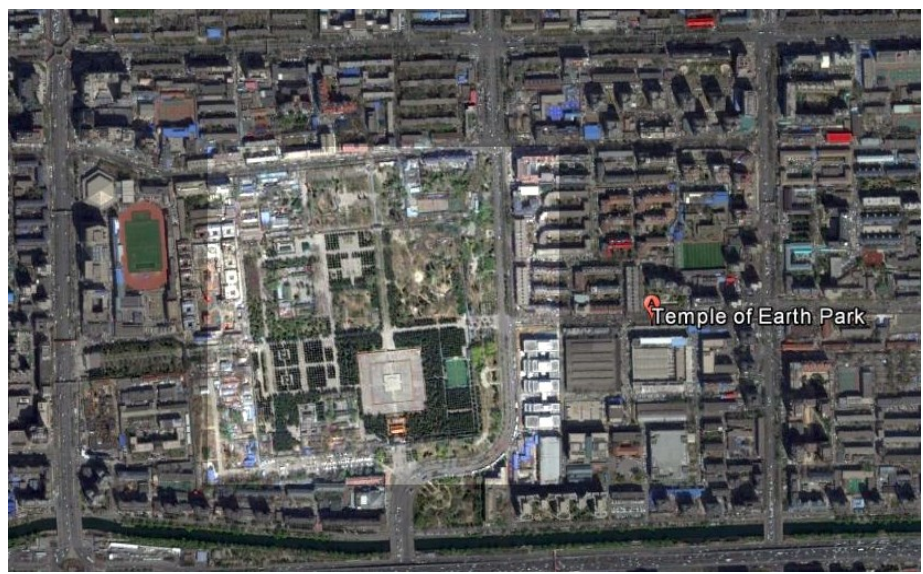

**Figure 6:** The Temple of the Earth in Beijing (Courtesy Google Earth).

The angle is so large that we can appreciate the dislocation of the complex with respect to the orthogonal layout of the streets. This Temple (also known as the Ditan Park) was built in 1530 by the Jiajing Emperor, during the Ming dynasty. Of the same year are the Temple of the Moon and the Temple of the Sun. In the Figure 7, we can see the altar of the temple of the Sun. We observe an angle of about 4.5 degrees, with respect to the direction of true North. For the Temple of the Moon (Figure 8), we have an angle of about 4 degrees. If these complexes of 1530 were surveyed using a magnetic compass, we could argue that the magnetic declination changed of about three degrees from 1420 to 1530. This conclusion needs further studies.

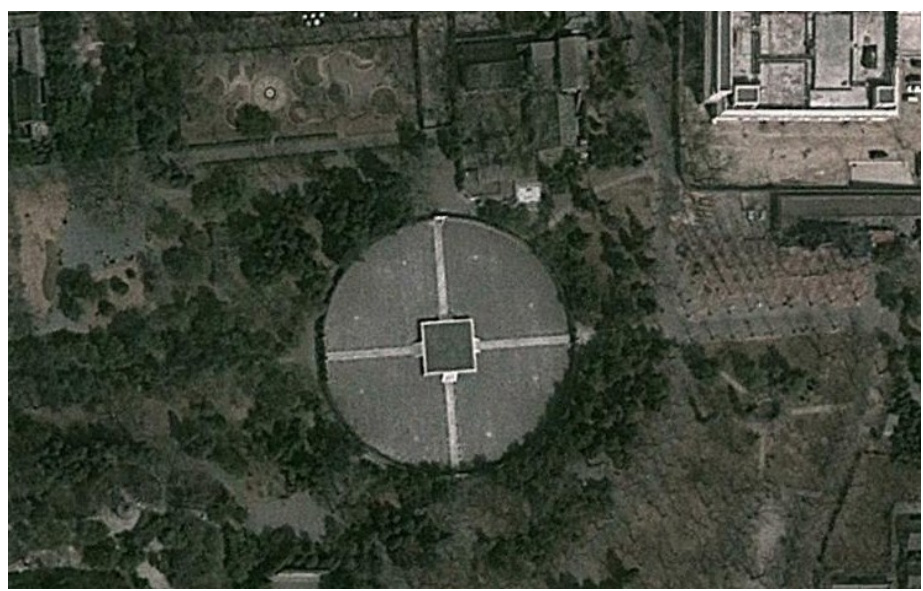

**Figure 7:** The altar of the Temple of the Sun (Courtesy Google Earth).

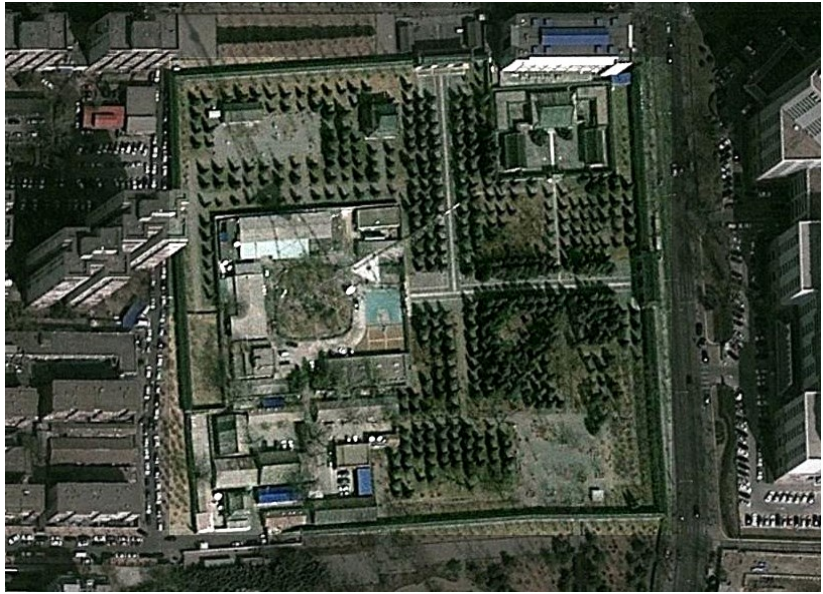

**Figure 8:** The Temple of the Moon (Courtesy Google Earth).

As we can see from the examples given above, the research is quite complex, however, it is rather interesting. We can try to compare the results from Figures 6-8, with the historical data on magnetic declination, given by [18,19]. These data are ranging from 1590 to 1990, as shown in the Figure 9. The line of zero declination is today far from China.

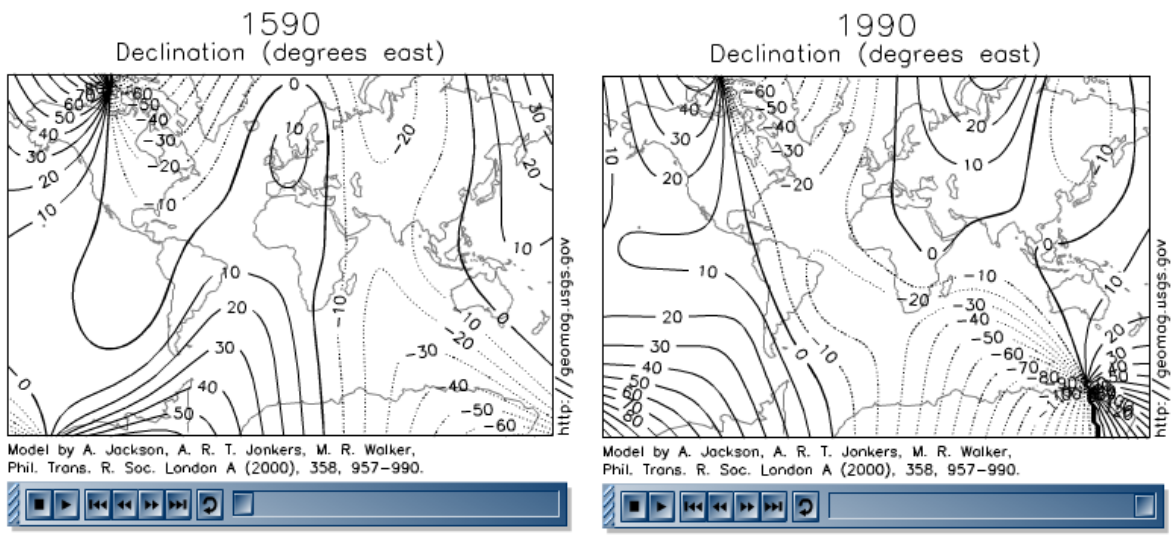

**Figure 9:** Thanks to the works of the USGS Geomagnetism Program [18], we have the possibility to see the evolution of the magnetic declination. Here we reproduce two frames of a movie, concerning the period from 1590 to 1990, made according to data in Ref.19.

We have the possibility to use also another remarkable interactive magnetic map. We can find it at the web site https://maps.ngdc.noaa.gov/viewers/historical_declination/. It is provided by the National Oceanic and Atmospheric Administration (NOAA) and developed from data of Ref.19, by means of model gufm1. As told by the site, we have four centuries of geomagnetic secular variation from historical records, giving us the "most complete picture of the evolution of the geomagnetic field, both at Earth's surface and at the core surface from 1590 to 1990".

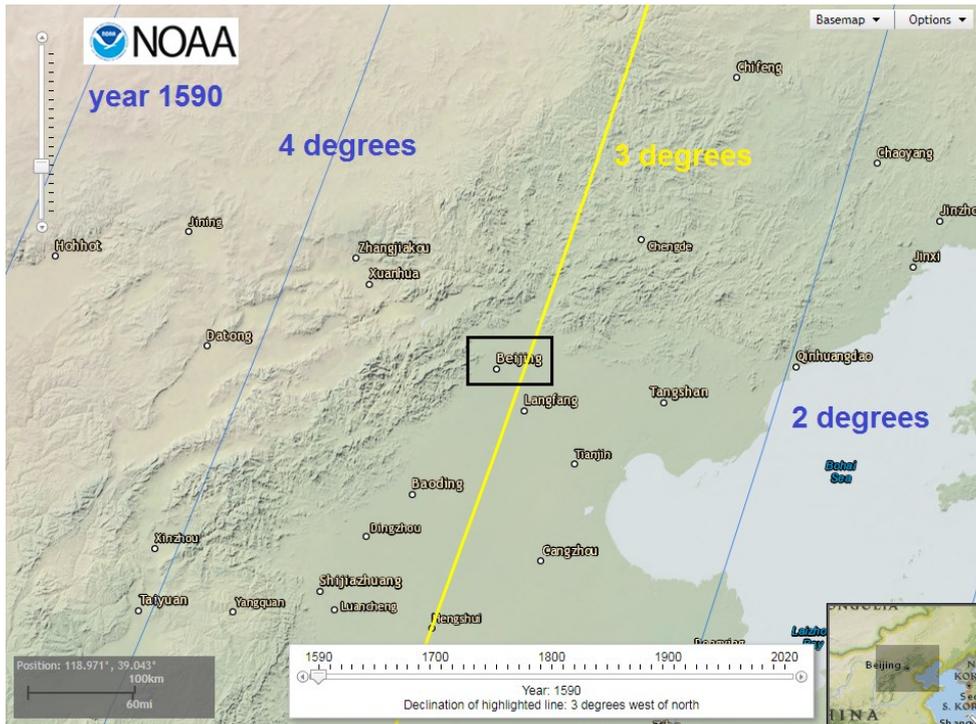
**Figure 10:** Thanks to the interactive map of NOAA, we can see that, on 1590, the magnetic declination in Beijing was between 3 and 4 degrees, negative.

Using the interactive map of NOAA we can estimate that, on 1590, the magnetic declination in Beijing was between 3 and 4 degrees, negative. This result is not so far from the angle of about 5 degrees that we have from the images of Figures 6-8. However, these architectonic complexes were built in 1530, and the Figure 10 is showing data of 1590. Therefore, let us consider the trend of the magnetic lines. For instance, in what direction was moving the line of the three degrees of declination? The answer is coming from the Figure 11.

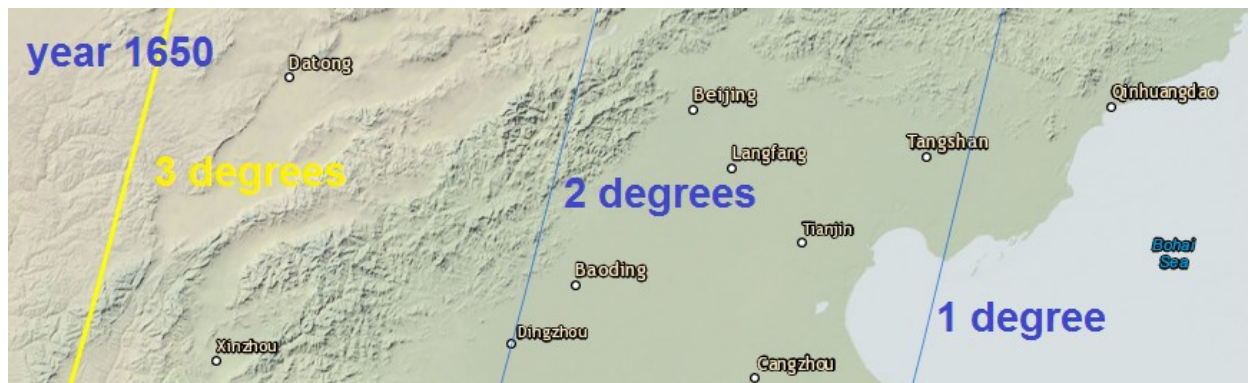
**Figure 11:** Again, using the interactive map of NOAA we can see that, on 1650, the magnetic declination in Beijing was between 1 and 2 degrees, negative.

The declination line of 3 degrees moved westward during the period from 1590 and 1650. Assuming the same trend in the period from 1530 to 1590, we can conclude that the magnetic declination in Beijing was, in 1530, of about 5 degrees (negative). This is in agreement with the architectonical results obtained from the Temple of the Earth and the Temples of the Sun and the Moon.